\documentclass[a4paper]{PoS}

\title{Spectral Methods in Causal Dynamical Triangulations}

\ShortTitle{Spectral Methods in Causal Dynamical Triangulations}

\author{\speaker{Giuseppe Clemente}, $^{ab}$ Massimo D'Elia $^{ab}$ and
    Alessandro Ferraro$^{ab}$\\
    \llap{$^a$} Universit\`a di Pisa, Largo B.~Pontecorvo 3, I-56127 Pisa, Italy\\
    \llap{$^b$} INFN Sezione di Pisa, Largo B.~Pontecorvo 3, I-56127 Pisa, Italy\\
        E-mail: \email{giuseppe.clemente@pi.infn.it},
    \email{massimo.delia@unipi.it}, \email{alessandro.ferraro@pi.infn.it}}

\abstract{We show recent results of the application of spectral analysis
in the setting of the Monte Carlo approach to Quantum Gravity known as
Causal Dynamical Triangulations (CDT), discussing the behavior of the
lowest lying eigenvalues of the Laplace-Beltrami operator computed on
spatial slices. This kind of analysis provides
information about running scales of the theory and about the critical
behaviour around a possible second order transition in the CDT phase
diagram, discussing the implications for the continuum limit.}

\FullConference{37th International Symposium on Lattice Field Theory - Lattice2019\\
		16-22 June 2019\\
		Wuhan, China}

\begin{document}

\section{Introduction}\label{sec:intro}


Formulating a consistent quantum field theory of gravity has proved to be a
difficult task,
since the action correctly describing the classical theory, the Einstein-Hilbert
action, is nonrenormalizable from a perturbative point of
view~\cite{sagnotti}.
%
However, there is the possibility of a nonperturbative mechanism for
renormalization, first envisaged by Weinberg and known as \emph{Asymptotic
safety scenario}~\cite{ass_weinberg}: the region of vanishing (bare) couplings in the phase
diagram is not necessarily an UV fixed point for the RG flow, but this does not
excludes that an UV fixed
point actually exists in another region of the phase diagram, not under the
reach of perturbative asymptotic series.\\
%
The RG flow of the Einstein-Hilbert theory, plus additional terms in the action,
can be studied nonperturbatively using Functional Renormalization Group techniques (FRG)~\cite{frg_qeg}. 
%
%
Another nonperturbative approach is numerical, based on Monte Carlo simulations
of a lattice-regularized field theory. The approach we discuss in the present
report is called Causal Dynamical Triangulations (CDT), where the action used is
the Einstein-Hilbert one, discretized in the Regge formalism~\cite{regge}, 
and with manifolds approximated as piecewise-linear manifolds with 
appropriate causal conditions.\\
%
%
The main goal of the CDT program is to show that, in the phase diagram 
of the theory, there exists a 
second order critical point where some definition of
correlation length diverges, and that the critical region around such point
describes a theory of quantum gravity consistent with our 
current understanding of the expected semiclassical behavior.\\
The CDT phase diagram shows a very rich structure with four phase identified
as $A$, $B$, $C_{dS}$ and $C_{b}$, and two possible second order lines; we refer the reader
to~\cite{cdt_report12,cdt_report19} for more details.
%

One of the most difficult problems in CDT without matter fields is the 
definition of local observables that can be associated to correlation lengths,
in order to properly investigate the critical 
properties of the system and the continuum limit.\\
Some recent advances in this direction came from the introduction of spectral
analysis of CDT configurations~\cite{cdt-LBseminal}, that is, the analysis of
eigenvalues and eigenvectors of the Laplace-Beltrami operator.
Here we discuss some of the results obtained, focusing on the analysis of the
critical properties of the (seemingly second order) transition line 
between phase $C_{b}$ and $C_{dS}$, which represents the most promising 
candidate for taking the continuum limit of a quantum theory of gravity.

\section{Numerical setup and spectral graph methods}\label{sec:numsetup}


Unlike the similar approach called Euclidean Dynamical
Triangulations (EDT)~\cite{edt1,edt2,edt3,dt_forcrand,dt_syracuse}, 
in CDT one enforces a causal condition of global
hyperbolicity~\cite{causconds}.
This implies that any configuration (simplicial manifold) can
be foliated into spatial slices with well defined euclidean time.
Due to this condition, and to the absence of singular points in the
simplicial manifold, the topology of spatial slices is constant through slices.
Here we consider simulations with $S^3$ topology of the
slices and with periodic conditions in the Euclidean time.

Spatial slices in CDT have only spatial links of uniform length 
(the lattice spacing); this allows us to faithfully represent
them by their \emph{dual graph}, that is the pair of sets $G=(N,E)$,
where the set of \emph{nodes} $N$ corresponds to 
spatial tetrahedra, while the set of \emph{edges} represents the adjacency
relations between tetrahedra connected by a facet.
The graph analogue of the \emph{Laplace-Beltrami} operator is called the 
\emph{Laplace matrix}, which is sparse, linear and
symmetric, and acts on the space of real-valued functions on $N$ ($\{f : N \rightarrow
\mathbb{R}\} \cong \mathbb{R}^{|N|}$); see reference~\cite{cdt-LBseminal} for details.
%

The updating algorithm used for sampling the path integral is
Metropolis-Hastings one, consisting of a set of local moves
(see Ref.~\cite{cdt_report12,cdt_report19} for more details).
In all simulations we used $N_t = 80$ spatial slices, keeping the
total spatial volume $V_{S,tot}$ fluctuating around a fixed target volume by
tuning one of the bare parameters ($k_4$).

In the following section we introduce some idea which we used to analyze and
interpret numerical results.

%
%
%
\subsection{Useful concepts from spectral analysis}\label{subsec:spectra-graphs}

The most interesting spectral quantities, describing the spatial geometries at large
scales, are the eigenvalues in the lower spectrum, in particular the lowest
(non-zero) one $\lambda_1$, which is called \emph{spectral gap} or usually
referred to as \emph{algebraic connectivity}, since it quantifies the degree of
connectedness of the manifold or graph (see ref.~\cite{cdt-LBseminal}).
Notice that the LB operator is the one which enters the diffusion equation (or
the wave equation), therefore, one can also interpret its eigenvalues
as the diffusion rates (or the wavenumber squared) of their associated eigenmodes:
a large value of the spectral gap means a fast diffusion (or small
wavelength) for the slower mode, which, in turn, means an overconnected
geometry, where all space is collapsed into a small region 
(the diameter is observed to grow logarithmically with the volume).
Another quantity of interest is $n(\lambda)$, which counts how many eigenvalues
fall below $\lambda$, which, for smooth manifolds, has an useful asymptotic behaviour called
\emph{Weyl's law}~\cite{weylslaw_1}:
\begin{equation}\label{eq:weyl_law}
    n(\lambda) \sim \frac{\omega_d}{{(2 \pi)}^d} V \lambda^{d/2},
\end{equation}
where $V$ and $d$ are respectively the volume and the local dimension of the
manifold, while $\omega_d$ is the volume of a $d$-dimensional ball of unit 
radius.
Inspired by Eq.~{(\ref{eq:weyl_law})}, we defined
a new function $d_{EFF}(\lambda)$, whose role is the one of an effective 
dimension of the manifold as a function of the length-scale
(dictated by $\lambda$):
\begin{equation}
    d_{EFF}(\lambda) \equiv 2 \frac{d \log (n/V)}{d \log \lambda}.
\label{eq:weyl_dimension} 
\end{equation}

\section{Numerical Results}\label{sec:results}

\begin{figure}
\centering
\begin{minipage}{.45\textwidth}
  \centering
  	\includegraphics[width=0.99\linewidth]{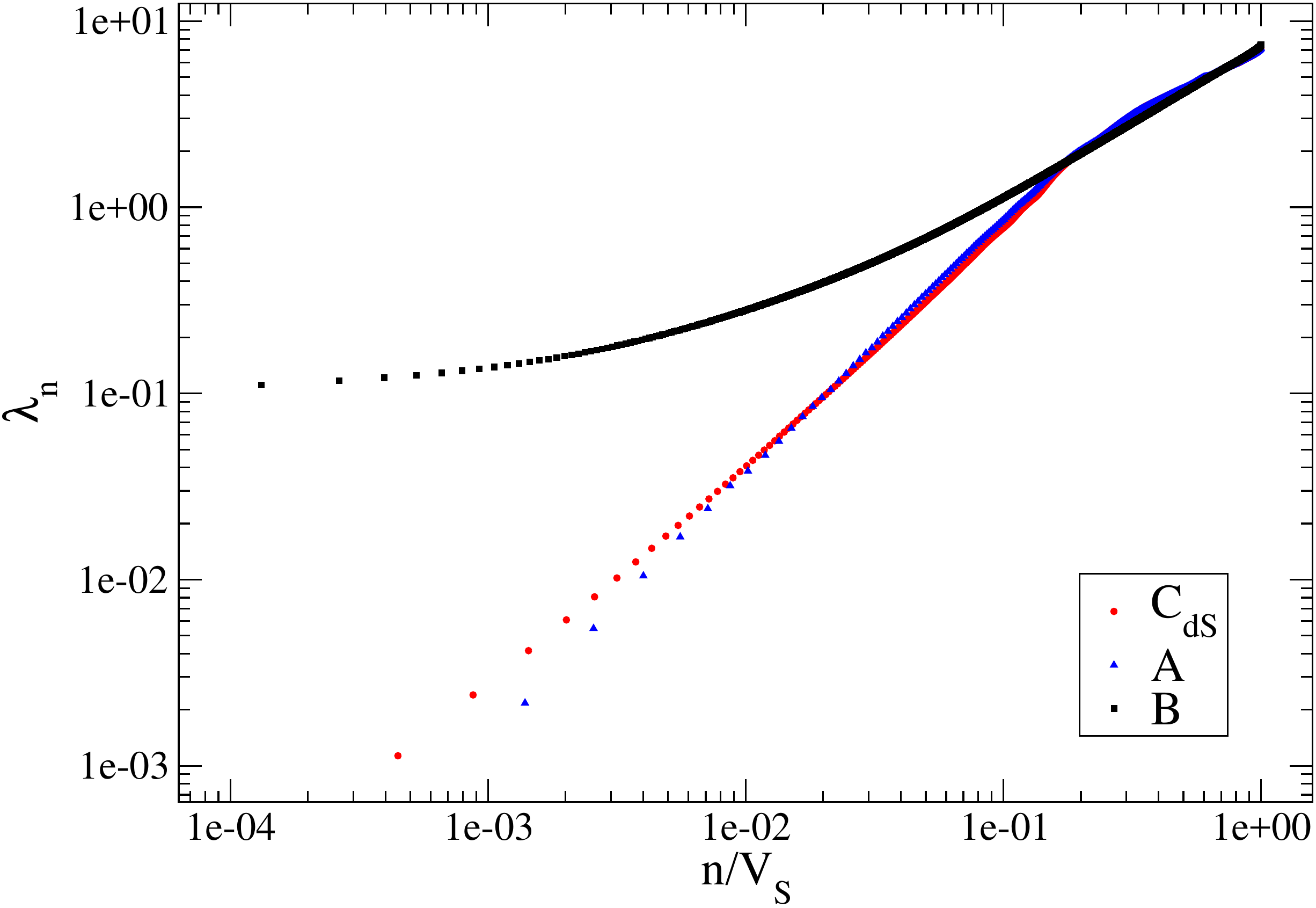}
    \caption{Averages of $\lambda_n$ versus $n/V_S$ computed in bins of $n/V_S$
    for slices taken from configurations sampled deep into the $A$, $B$ and
    $C_{dS}$ phases: $(k_0,\Delta) = (5,0.6)$, $(2.2,-0.2)$ and $(2.2,0.6)$ respectively.
The volume is fixed to $V_{S,tot}=40k$ for configurations in $A$ and $C_{dS}$ phase,
and to $V_{S,tot}=8k$ for configurations in $B$ phase.}\label{fig:averbin_CdS_A_B-lam-vs-k_V}
\end{minipage}
\hspace{1em}
\begin{minipage}{.45\textwidth}
  	\centering
  	\includegraphics[width=0.99\linewidth]{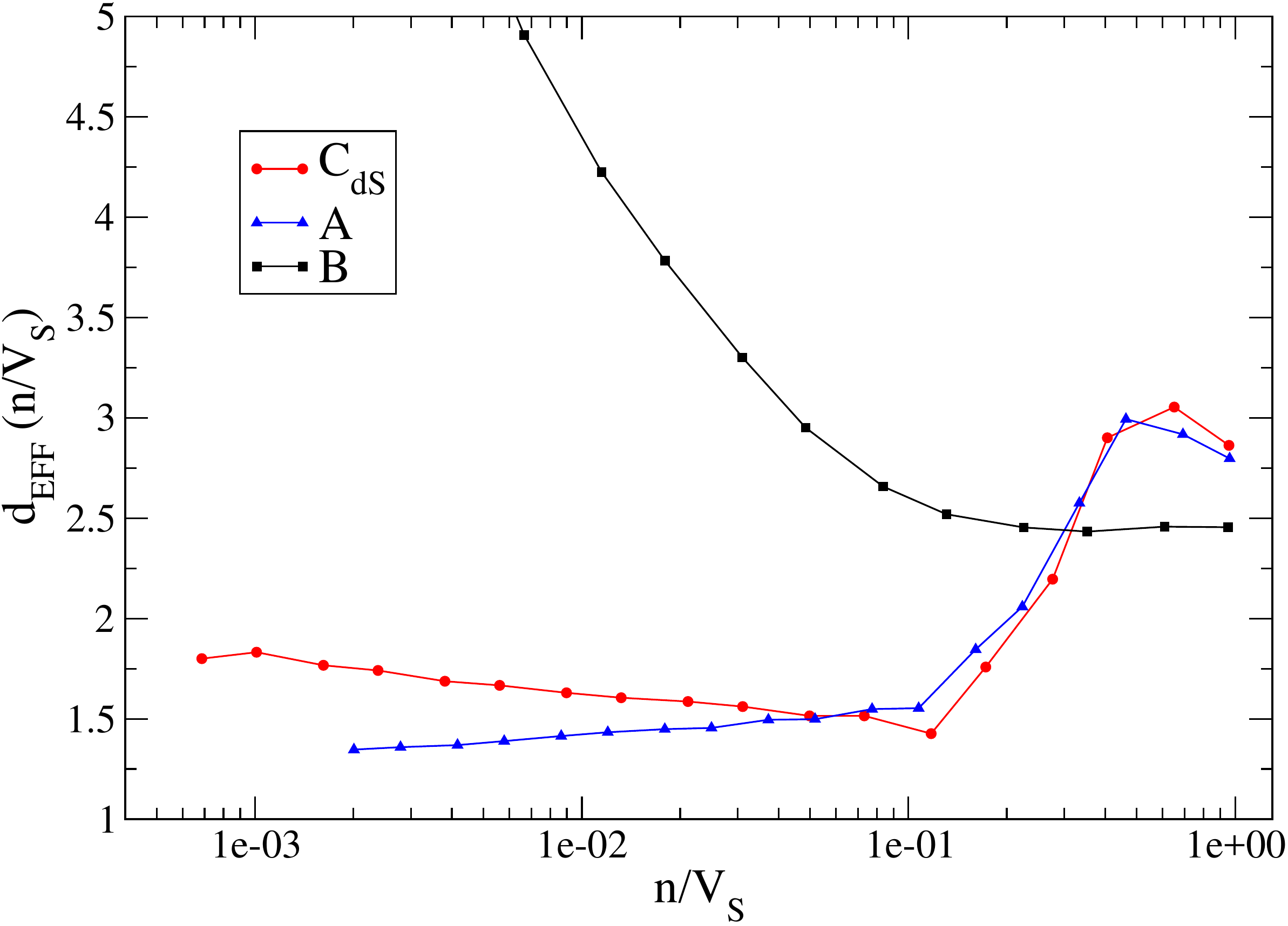}
    \caption{Running effective dimension $d_{EFF}$ (using definition Weyl's law) 
        for the curves shown on the left Figure, obtained by running averages of
        the logarithmic derivative.
        The curve associated to the $B$ phase is diverging for $n/V_S \rightarrow 0$,
        but we focused on the range $d_{EFF}\in\left[1,5\right]$ for readability of the other two curves.}
        \label{fig:plot_CdS_A_B-Dk_V}
\end{minipage}
\end{figure}

We begin by considering the spectra of the Laplace matrix for spatial slices of 
configurations deep into the $A$, $B$ and $C_{dS}$.\\
Eqs.~{(\ref{eq:weyl_law})} and
{(\ref{eq:weyl_dimension})} suggest the possibility of an universal scaling of
$n/V_S$ as a function of $\lambda$: indeed we observe a remarkable collapse in
the scatter plots $\lambda$ vs $n/V_S$, independent of the slice (spatial) 
volume $V_S$, but dependent, in general, on the position in the phase diagram.
Figure~\ref{fig:averbin_CdS_A_B-lam-vs-k_V} shows the averages of $\lambda$ in
bins of $n/V_S$, for the three phases aforementioned; these curves are then
used to estimate the running effective dimension $d_{EFF}$ by 
Eq.~{(\ref{eq:weyl_dimension})}, as shown in
Figure~\ref{fig:plot_CdS_A_B-Dk_V}.\\
The main distinguishing feature of spectra of slices in phase $B$ (actually its
unique spatial slice for each configuration) is the presence of a large spectral
gap $\lambda_1$, which is observed to be independent on the total volume.
As argued in the previous section, this large value is associated to a 
highly-connected geometry: indeed, the diverging effective dimension at larger
scales (small $n/V_S$ from Weyl's correspondence) suggests that slices in $B$
phase do not seem similar to manifolds at all.
The $C_{dS}$ and $A$ phases, even if differing by the correlation between slices
with near slice-times (high for
$C_{dS}$, but vanishing for $A$), show a similar behavior of the $\lambda$ vs
$n/V_S$ distributions. In particular, we observe a vanishing spectral gap, with a
power-like scaling at the large scales; the running effective dimension observed
stays relatively constant in the large to intermediate scale, but here it is 
fractional, showing a fractal-like behavior of the geometry.\\

\subsection{Spectral properties of the $C_b$ phase}
\begin{figure}
\centering
\begin{minipage}{.45\textwidth}
  \centering
		\includegraphics[width=0.99\linewidth]{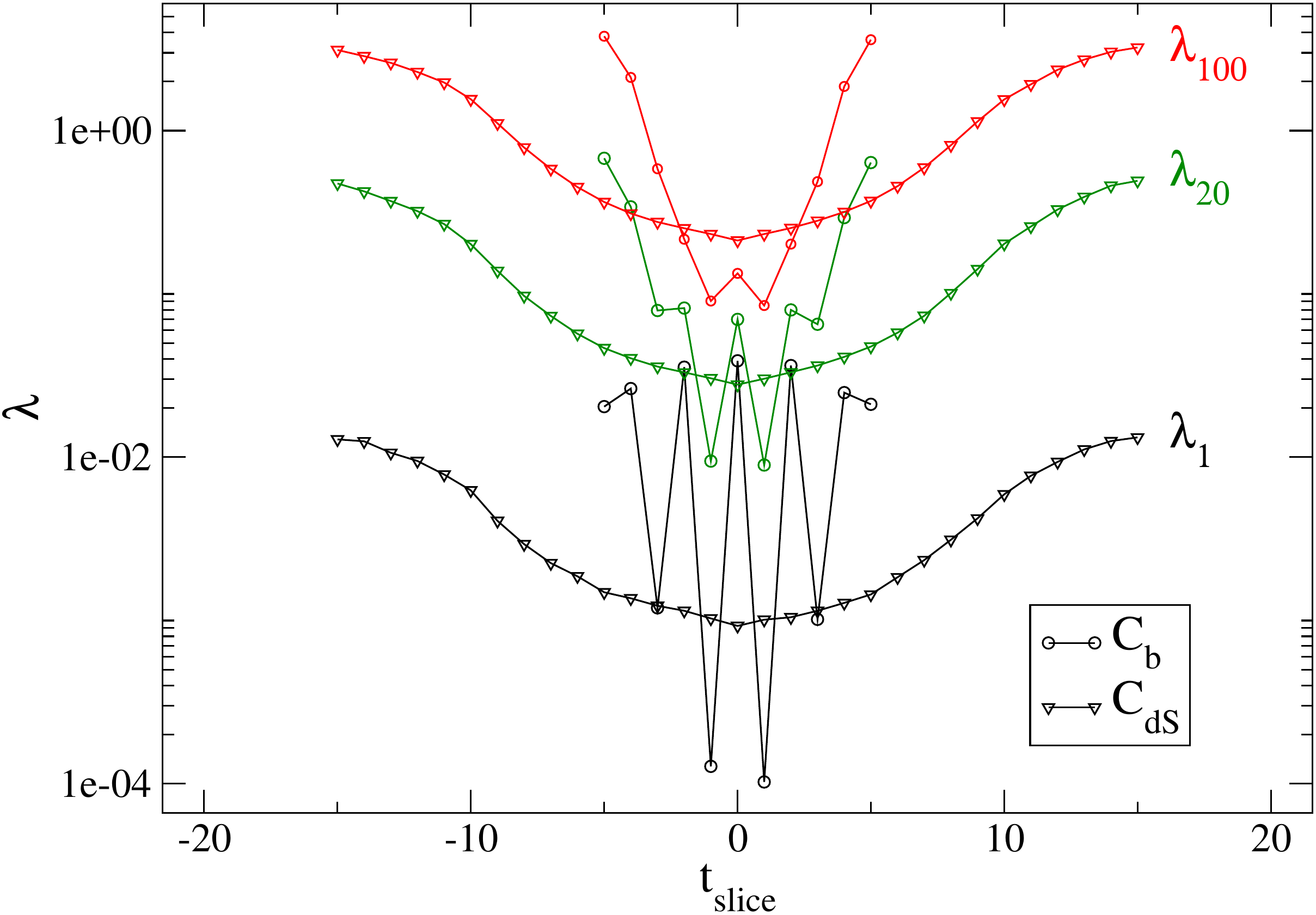}
		\caption{Averages of $\lambda_1$, $\lambda_{20}$ and $\lambda_{100}$ as a 
		function of the slice-time for configurations in $C_b$ and $C_{dS}$
    phases.}\label{fig:aver_Cb-CdS_lam1-20-100_vs_tslice}
\end{minipage}
\hspace{1em}
\begin{minipage}{.45\textwidth}
		\centering
    \includegraphics[width=0.95\linewidth]{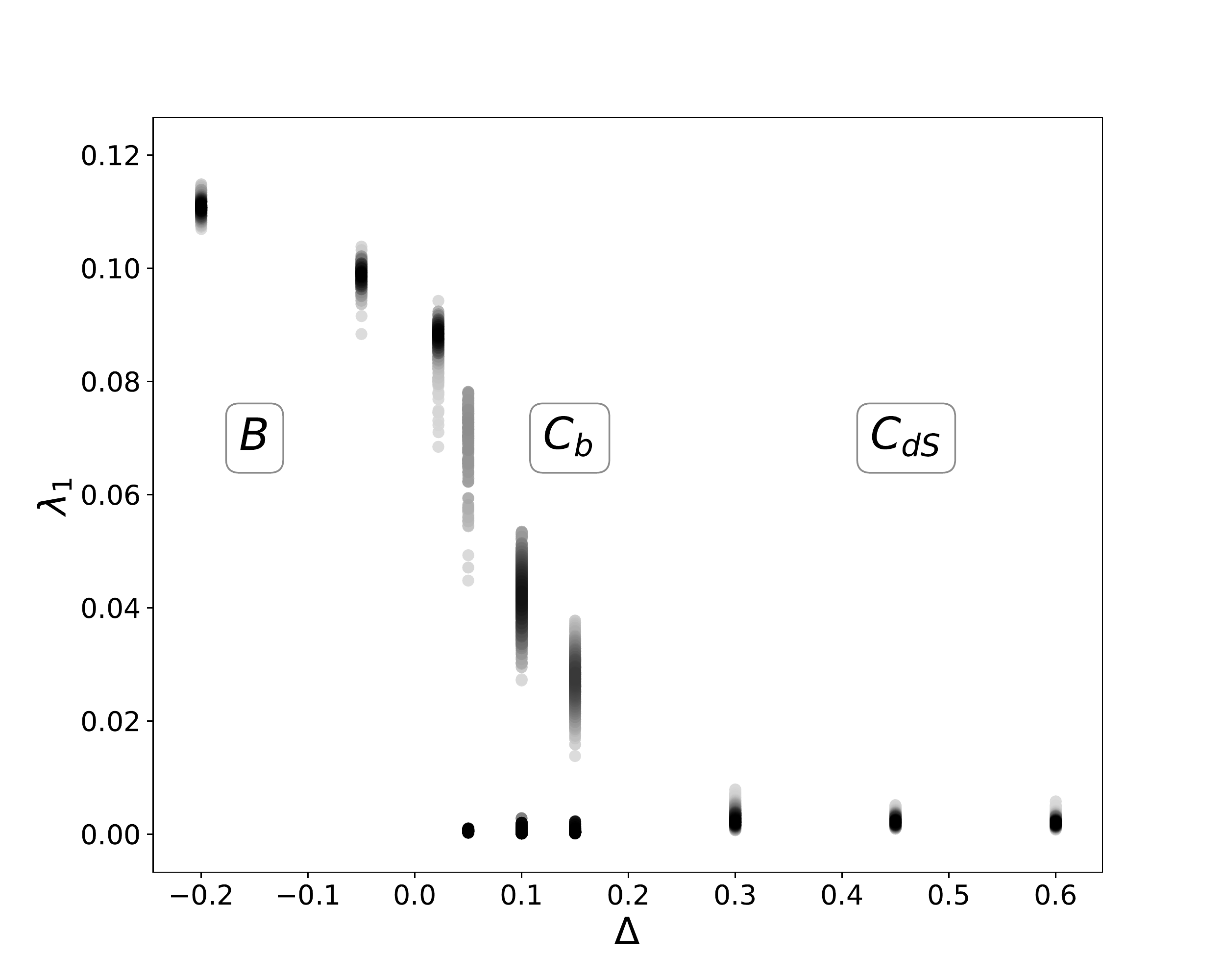}
    \caption{Distributions of $\lambda_1$ (vertically) for $k_0=2.2$ and variable $\Delta$ for configurations with total spatial volume $V_{S,tot}=\frac{N_{41}}{2}=40k$}\label{fig:lambda1_op-scatt}
\end{minipage}
\end{figure}

Configurations in the $C_b$ phase have the characteristic feature that their
slices have geometric properties which alternate in the
slice-time~\cite{cdt_charnewphase,cdt_newhightrans} so that two distinct classes of slices can
be identified. This fact is evident in the low-lying spectra, 
as can be seen in
Figure~\ref{fig:aver_Cb-CdS_lam1-20-100_vs_tslice}:
slices in the bulk of 
$C_b$ phase configurations can be separated in two distinct classes by the value
of their spectral gap, changing abruptly for adjacent slices even by two orders
of magnitude; in the $C_{dS}$ phase, instead, there is no sharp distinction,
apart from a volume-dependent behavior connected to Weyl-like scaling.
Indeed, the geometric properties of the two classes of slices parallel the
properties of the $B$ phase (for the gapped ones) and of the $C_{dS}$ phase (for the
non-gapped ones), which brought us to call them respectively $B$-like and
$dS$-like.
A more general view of the distribution of spectral gaps for different points of
the phase diagram (in a fixed $k_0=2.2$ line) is given in
Figure~\ref{fig:lambda1_op-scatt}, which shows well the separation of in the two
classes for $C_b$ slices, and how this separation disappears in going from the
$C_{b}$ to the $C_{dS}$ phase.
This observation suggests to use the spectral gap of $B$-like slices in the
$C_{b}$ phase as an order parameter for the $C_{b}$-$C_{dS}$ transition, which in
literature has been claimed to be second-order by using different
methods~\cite{cdt_charnewphase,cdt_newhightrans}.
This will be the focus of the next section.

\subsection{Spectral study of the critical properties of the $C_{b}$-$C_{dS}$ transition}

\begin{figure}
\centering
\begin{minipage}{.45\textwidth}
		\centering
    \includegraphics[width=0.95\columnwidth, clip]{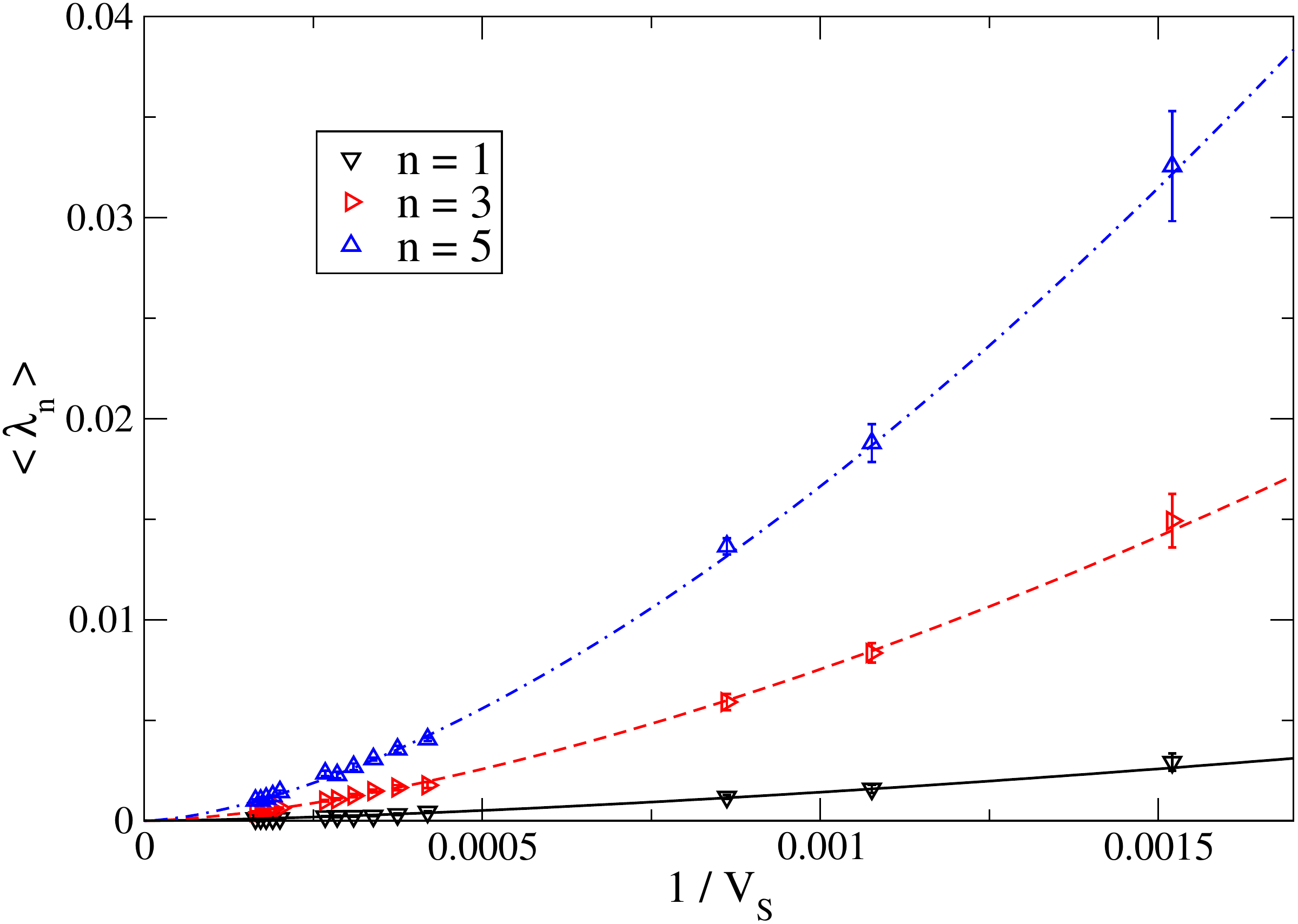}
    \caption{Weyl scaling for some eigenvalue $\langle\lambda_n\rangle$ ($n=1,3,5$) of
    $dS$-like slices at $(k_0,\Delta)=(0.75,0.4)$, with vanishing extrapolations in the thermodynamical limit.}\label{fig:cb_ds_scaling}
\end{minipage}
\hspace{1em}
\begin{minipage}{.45\textwidth}
		\centering
    \includegraphics[width=0.95\columnwidth, clip]{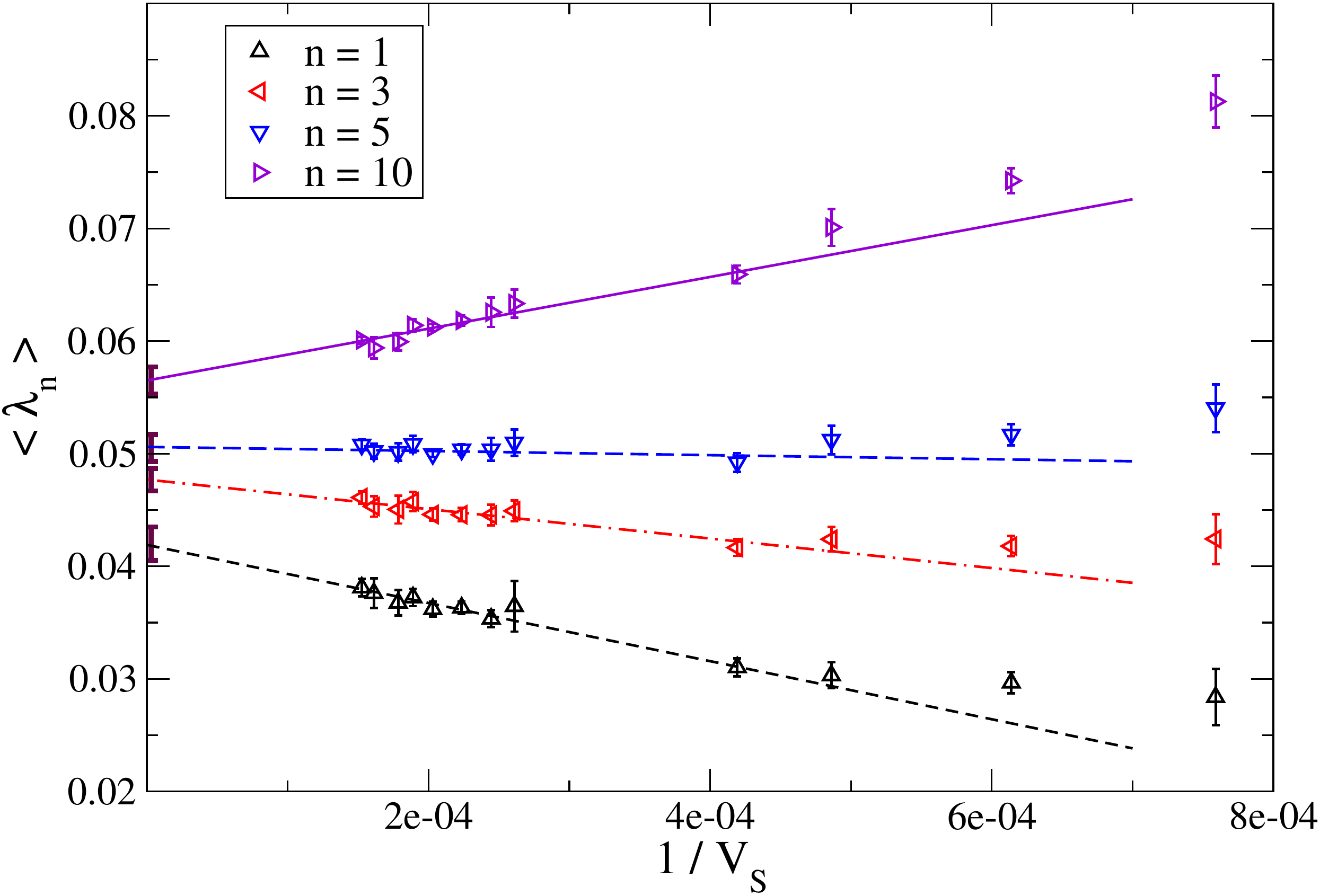}
    \caption{Weyl scaling for some eigenvalue $\langle\lambda_n\rangle$ ($n=1,3,5$) of
    $B$-like slices at $(k_0,\Delta)=(0.75,0.4)$, where the thermodynamical limit
extrapolations are shown on the vertical axis.}\label{fig:cb_b_scaling}
\end{minipage}
\end{figure}

First of all, in order to use the spectral gap of $B$-like slices in $C_b$
configurations, we need to study separately the two classes of slices (details
can be found in~\cite{cdt-LBrunningscales}).
Figure~\ref{fig:cb_ds_scaling} shows a plot of $\langle\lambda_n\rangle$ vs $1/V_S$ for the
$dS$-like class, where the behavior is compatible
($\chi^2/$d.o.f. $\simeq 1$) with a vanishing gap in the thermodynamical
limit (as expected by Weyl's law). The picture is different for $B$-like slices
(Figure~\ref{fig:cb_b_scaling}),
where the extrapolations to thermodynamical limit show, not only a non-vanishing
spectral gap $\langle\lambda_1\rangle_\infty$, but distinct values of the
extrapolations for different eigenvalues $\langle\lambda_n\rangle_\infty$; this
fact will be discussed in the conclusions.\\

\begin{figure}[t]
    \centering
    \includegraphics[width=0.75\columnwidth, clip]{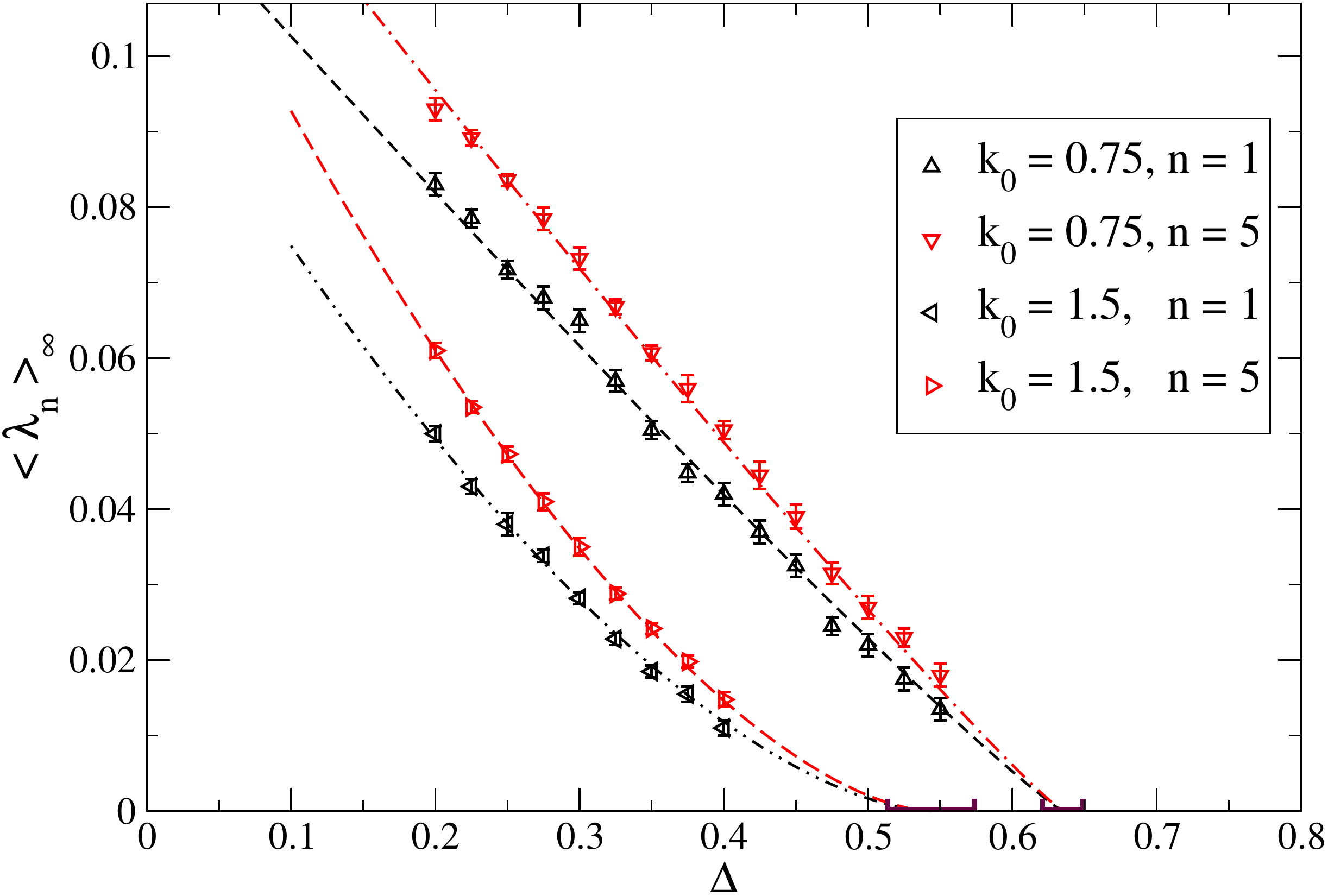}
    \caption{$\langle \lambda_n \rangle_\infty$ 
in $B$-like slices as a function of $\Delta$ for different
values of $k_0$ and $n$, together with best fits of the scaling equation. 
    }\label{fig:bicritical}
\end{figure}

In Figure~\ref{fig:bicritical} we report
some $\langle \lambda_n \rangle_\infty$ 
($n = 1$ and $5$)
as a function of $\Delta$ along two lines of simulations at $k_0=0.75$ and
$k_0=1.5$; since, on the immediate vicinity of the critical point, $B$-like and
$dS$ like classes mix, it is difficult to separate them and take the
thermodynamical limit, therefore, some values 
of $\Delta$ are excluded from Figure~\ref{fig:bicritical}.

On dimensional grounds, different $\langle \lambda_n \rangle_\infty$ 
are associated to different 
inverse squared lengths, which however, for the continuum limit to be
meaningful,
should scale proportionally to each other around a critical point, i.e., with a
common critical index.
Based on this observation, we tried to fit the data using the scaling formula
\begin{equation}
    \langle \lambda_n \rangle_\infty  = A_n {(\Delta_c - \Delta)}^{2 \nu},
\label{crit_fit}
\end{equation}
where only the $A_n$ coefficients depend on $n$.
A combined fit, including $n = 1, 5$, yields
$\Delta_c = 0.635(14)$, $\nu = 0.55(4)$ for 
$k_0 = 0.75$ ($\chi^2/{\rm d.o.f.} = 31/26$), and 
$\Delta_c = 0.544(36)$, $\nu = 0.82(12)$ for 
$k_0 = 1.50$ ($\chi^2/{\rm d.o.f.} = 6/14$).
Similar and consistent results are obtained including 
different values of $n$, or if the eigenvalues
are fitted separately.
In principle, our best fits suggest that the index $\nu$ may change along
the transition line; however, a global
fit in which $\nu$ is set equal for both $k_0$ works
equally well, yielding $\nu = 0.59(4)$, 
$\Delta_c (k_0 = 0.75) = 0.656(15)$, $\Delta_c (k_0 = 1.5) = 0.479(10)$
with $\chi^2/{\rm d.o.f.} = 47/41$.

\section{Conclusions}\label{sec:conclusions}


%
%
%

We discussed some recent results obtained from the analysis of the spectrum of
the LB operator define on spatial slices of CDT
configurations~\cite{cdt-LBseminal,cdt-LBrunningscales}, focusing in
particular on the role of the $C_b$ phase and the $C_b$-$C_{dS}$ transition,
which represents the most promising candidate for a continuum limit. In
particular, we have shown that the two classes of slices in the $C_b$ phase 
have geometric characteristics analogue to the $B$ and $C_{dS}$ phases, with a
gapped and non-gapped structure of the spectra respectively.\\
The thermodynamical limit of the lowest eigenvalues $\langle \lambda_n
\rangle_\infty$ of $B$-like slices shows the emergence of a hierarchy of
distinct characteristic lengths, which could be interpreted as a set of
dynamical mass scales, connected to the propagation of massless scalar
fields in the sampled geometries. Approaching the $C_{dS}$ phase, these mass
scales appear to vanish with a common critical index, as expected by a
lattice field theory approaching the continuum limit. 



\begin{thebibliography}{99}
		\bibitem{sagnotti}
		M.~H.~Goroff, A.~Sagnotti,
        \emph{The Ultraviolet Behavior of Einstein Gravity},
        \emph{Nucl.\ Phys.\ B} {\bf 266} (1986) 709.

		\bibitem{ass_weinberg}
		S.~Weinberg,
        \emph{General Relativity, an Einstein Centenary Survey}, ch.16
	    Cambridge Univ. Press, 1979.
		
		
%
%
		
		\bibitem{frg_qeg}
		M.~Reuter and F.~Saueressig,
        \emph{Quantum Einstein Gravity},
        \emph{New J.\ Phys.}\  {\bf 14} (2012) 055022
        [{\tt arXiv:1202.2274 [hep-th]}].
		
%
		
		
		\bibitem{regge} 
		T.~Regge,
        \emph{General Relativity Without Coordinates},
        \emph{Nuovo Cim.}\  {\bf 19}, 558 (1961).

		\bibitem{cdt_report12} 
		J.~Ambjorn, A.~Goerlich, J.~Jurkiewicz and R.~Loll,
        \emph{Nonperturbative Quantum Gravity},
        \emph{Phys.\ Rept.}\  {\bf 519}, 127 (2012)
        [{\tt arXiv:1203.3591 [hep-th]}].

        \bibitem{cdt_report19}
          R.~Loll,
          \emph{Quantum Gravity from Causal Dynamical Triangulations: A Review},
          [{\tt arXiv:1905.08669 [hep-th]}].


        \bibitem{cdt-LBseminal}
        G.~Clemente and M.~D'Elia,
        \emph{Spectrum of the Laplace-Beltrami operator and the phase structure
        of causal dynamical triangulations},
        \emph{Phys.\ Rev.\ D} {\bf 97} (2018) no.12, 124022
        doi:10.1103/PhysRevD.97.124022
        [{\tt arXiv:1804.02294 [hep-th]}].
		
		\bibitem{edt1} 
		J.~Ambjorn and J.~Jurkiewicz,
        \emph{Four-dimensional simplicial quantum gravity},
        \emph{Phys.\ Lett.\ B} {\bf 278}, 42 (1992).


		\bibitem{edt2} 
		M.~E.~Agishtein and A.~A.~Migdal,
        \emph{Simulations of four-dimensional simplicial quantum gravity},
        \emph{Mod.\ Phys.\ Lett.\ A} {\bf 7}, 1039 (1992).
		
		\bibitem{edt3}
		J.~Ambjorn, J.~Jurkiewicz and C.~F.~Kristjansen,
        \emph{Quantum gravity},
        \emph{Nucl.\ Phys.\ B} {\bf 393} (1993) 601
        [{\tt hep-th/9208032}].
		
		\bibitem{dt_forcrand}
		T.~Rindlisbacher and P.~de Forcrand,
        \emph{Euclidean Dynamical Triangulation revisited: is the phase
        transition really 1st order? (extended version)},
        \emph{JHEP} {\bf 1505} (2015) 138
        [{\tt arXiv:1503.03706 [hep-lat]}].
		
		\bibitem{dt_syracuse}
		J.~Laiho, S.~Bassler, D.~Coumbe, D.~Du and J.~T.~Neelakanta,
        \emph{Lattice Quantum Gravity and Asymptotic Safety},
        \emph{Phys.\ Rev.\ D} {\bf 96} (2017) no.6,  064015
        [{\tt arXiv:1604.02745 [hep-th]}].
		
		\bibitem{causconds}
		E.~Minguzzi and M.~Sanchez,
        \emph{The Causal hierarchy of spacetimes},
        \emph{EMS Pub.House}, 2008, p.299-358
        [{\tt gr-qc/0609119}].

         \bibitem{weylslaw_1}
        H.~Weyl,
        \emph{\"Uber die asymptotische Verteilung der Eigenwerte"},
        \emph{Nachr. Konigl. Ges. Wiss. G\"ottingen}, 110--117 (1911). 

%
%
%
		\bibitem{cdt_charnewphase}
		J.~Ambjorn, J.~Gizbert-Studnicki, A.~Goerlich, J.~Jurkiewicz, N.~Klitgaard and R.~Loll,
        \emph{Characteristics of the new phase in CDT},
        \emph{Eur.\ Phys.\ J.\ C} {\bf 77} (2017) no.3,  152
        [{\tt arXiv:1610.05245 [hep-th]}].
		
		
		\bibitem{cdt_newhightrans}
		J.~Ambjorn, D.~Coumbe, J.~Gizbert-Studnicki, A.~Goerlich and J.~Jurkiewicz,
        \emph{New higher-order transition in causal dynamical triangulations},
        \emph{Phys.\ Rev.\ D} {\bf 95} (2017) no.12,  124029
        [{\tt arXiv:1704.04373 [hep-lat]}].

         
		
%
%
		

%
%
%
%
%
%
%
%
%
%
%
%
%
%
%
%

        \bibitem{cdt-LBrunningscales}
        G.~Clemente, M.~D'Elia and A.~Ferraro,
        \emph{Running scales in causal dynamical triangulations},
        \emph{Phys.\ Rev.\ D} {\bf 99} (2019) no.11, 114506
        doi:10.1103/PhysRevD.99.114506
        [{\tt arXiv:1903.00430 [hep-th]}].


%
%
%
%
%
%
%
%

\end{thebibliography}
\end{document}